# Disordered proteins and network disorder in network descriptions of protein structure, dynamics and function. Hypotheses and a comprehensive review

Peter Csermely[1,*], Kuljeet Singh Sandhu[2], Eszter Hazai[3], Zsolt Hoksza[1,#], Huba J.M. Kiss[1,4], Federico Miozzo[5]; Dániel V. Veres[1]; Francesco Piazza[6] and Ruth Nussinov[7,8]

[1]Department of Medical Chemistry, Semmelweis University, Budapest, Hungary; [2]Genome Institute of Singapore; [3]Virtua-Drug Ltd., Budapest, Hungary; [4]Department of Ophthalmology, Semmelweis University, Budapest, Hungary; [5]UMR7216 Epigénétique et Destin Cellulaire, Université Paris Diderot, Paris, France; [6]Centre de Biophysique Moleculaire, CNRS, Rue Charles Sadron, 45071, Orléans, France; [7]Center for Cancer Research Nanobiology Program, SAIC-Frederick, Inc., NCI-Frederick, Frederick, MD 21702, USA and [8]Sackler Institute of Molecular Medicine, Department of Human Genetics and Molecular Medicine, Sackler School of Medicine, Tel Aviv University, Tel Aviv 69978, Israel

**Abstract:** During the last decade, network approaches became a powerful tool to describe protein structure and dynamics. Here we review the links between disordered proteins and the associated networks, and describe the consequences of local, mesoscopic and global network disorder on changes in protein structure and dynamics. We introduce a new classification of protein networks into 'cumulus-type', i.e., those similar to puffy (white) clouds, and 'stratus-type', i.e., those similar to flat, dense (dark) low-lying clouds, and relate these network types to protein disorder dynamics and to differences in energy transmission processes. In the first class, there is limited overlap between the modules, which implies higher rigidity of the individual units; there the conformational changes can be described by an 'energy transfer' mechanism. In the second class, the topology presents a compact structure with significant overlap between the modules; there the conformational changes can be described by 'multi-trajectories'; that is, multiple highly populated pathways. We further propose that disordered protein regions evolved to help other protein segments reach 'rarely visited' but functionally-related states. We also show the role of disorder in 'spatial games' of amino acids; highlight the effects of intrinsically disordered proteins (IDPs) on cellular networks and list some possible studies linking protein disorder and protein structure networks.

**Key words:** assortativity; clustering; entropy; interactome; intrinsically disordered proteins; signaling; soliton; symmetry

**Running title:** Disorder of proteins and networks

[*]Address all correspondence to this author at the Department of Medical Chemistry, Semmelweis University, H-1444 Budapest, P. O. Box 260, Hungary; Tel: +36-1-459-1500; Fax: +36-1-266-3802; E-mail: csermely@eok.sote.hu
[#]Mr. Zsolt Hoksza is a high school student of the Fazekas High School (www.fazekas.hu) in Budapest, Hungary, who started his research as a member of the EU Descartes Award winning High School Research Student organization (www.kutdiak.hu) founded by P.C. and offering research opportunities for several thousands of high school students since 1996.

## 1. INTRODUCTION

The concept of networks is compelling and powerful. Here, we analyze protein disorder in terms of the associated networks, and *vice versa*, disorder in the networks as an indicator of certain features of protein dynamics. In the following comprehensive review we will first summarize how a network description may help our understanding of protein structure, dynamics and function. In Sections 4 and 5 the consequences of disorder in protein structure and dynamics will be described in terms of their network representations. As a novel element of our review, we will summarize the current knowledge of disorder in network structures using both *general* terms, such as network entropy, and more detailed concepts at the local, mesoscopic and global levels of the network structure. The concept of network disorder will be extended to the dynamics of protein structure networks. In this Section the propagation of perturbations, the synchrony of oscillations and the role of network segments having specific dynamics will be discussed. We will show how network disorder reflects the structure and dynamics of proteins. In Section 8 we will introduce novel concepts. The key element of these novel ideas is the classification of proteins into 'cumulus-type networks' and 'stratus-type networks', where the differences between these two protein classes reflect the mechanisms of conformational changes. We will conclude the review by listing possible consequences of molecular disorder at higher levels of the cellular network hierarchy, such as in protein-protein interaction and signaling networks, and by highlighting future research possibilities in this direction.

## 2. STRUCTURE AND DYNAMICS OF PROTEINS AS NETWORKS

**2.A. Definition of protein structure networks**. Protein structure networks (also called protein contact networks) form the basal layer of the cellular network hierarchy. At the residue level of coarse-graining, the nodes are amino acids, while links connect amino acids whose inter-distance is below a cut-off (usually 0.4 to 0.85 nm) in the native fold. Protein structure networks may have weighted links instead of distance cut-offs, and may discriminate between individual atoms (like $\alpha C$ or $\beta C$ atoms). Covalent bonds may be included or excluded in the network representation [1-9].

**2.B. Key topological properties of protein structure networks.** In the small world of protein structure networks any two amino acids are separated by only a few links, and the network diameter grows logarithmically with increasing number of amino acids. This small-worldness promotes the fast transmission of perturbations (conformational changes). Typically, protein structure networks have a Poisson degree distribution rather than the widely characteristic scale-free degree distribution. This means that they have fewer hubs than expected, and amino acid "mega-hubs", having an extremely large number of neighbors do not exist, which is also expected. However, the existing (smaller) hubs play a major structural role. As an example, hubs were shown to increase the thermodynamic stability of proteins [3, 6, 8, 10-14].

Protein structure networks have modules (i.e. network communities), which are often hierarchical, and usually correspond to domains. Proteins having less than 200 amino acids seldom display a modular structure [5, 7, 13, 15-18], however they consist of 'foldons' [19], which may form a distinct class of tighter structural networks, with fast folding kinetics.

**2.C. Network description of protein dynamics: elastic network models.** Besides protein structure networks, other network-related methods, such as statistical-mechanics approaches of Gaussian network models, elastic network models and normal-mode analysis, are also useful to explore protein dynamics. Here, a harmonic potential is used to account for pair-wise interactions among either Cα atoms only (coarse-graining at the residue level) or among all atoms, forming a spring network [11, 20-28]. The low frequency collective modes (but not the medium to high frequency ones) are robust to network construction details once the essential contacts of the inner-most coordination shells have been included [29]. Combination of the elastic network model with protein structure networks showed that functionally active residues have enhanced communication properties [30].

**2.D. Other network descriptions of proteins.** Protein dynamics is usually described at the fastest (local, atomic) level by the ensemble of atomic vibrations or oscillations. Dynamics at the slower (mesoscopic) level describe conformational changes. All possible conformations of a given protein are summarized by energy networks, or conformational networks, where nodes are the individual conformations the protein may visit, and links represent the allowed transitions between these conformations (see the following reviews: [4, 28, 31, 32]). 'Protein dynamics' at the slowest (global) level can be assessed by mutations affecting the structures on evolutionary timescales (Section 8.E.). Here nodes are the individual forms of homologous proteins, and links show their evolutionary relatedness (usually in the form of structural similarity). The resulting network is often called the 'protein universe' (see e.g. [33]). We will mainly focus on protein structure networks with some consequences for the elastic network models. Conformational and evolutionary networks will not be addressed, since disorder has not been explored yet in these representations.

**3. PROTEIN STRUCTURE DISORDER IN NETWORKS**

Disorder of protein structures may be conceived at three levels of complexity: local (Section 3.A.), mesoscopic (Section 3.B.) and global (Section 3.C.). In the following Section we will consider local disorder through the most flexible and rigid amino acids, glycine and proline, respectively, as well as through flexible loops and intrinsically disordered segments. Disorder at the intermediate mesoscopic level will be introduced in the form of molten globules and intrinsically disordered domains. Finally, structural disorder of the whole protein will be described through examples of intrinsically disordered and unfolded proteins (see example in Section 4).

**3.A. Local disorder of protein structures.** A frequent example of local disorder arises from amino acids (or sequences) having a deviant local flexibility/rigidity pattern, like glycine or proline. Regretfully, no systematic studies analyzed the position of either glycines or prolines in protein structure networks so far. Loops also frequently determine local disorder, with sparse hydrogen bonds being either the cause or the outcome of this flexibility. Loops are important in folding and conformational transitions. In agreement with this, loops were shown to provide a major contribution to the slowest, most cooperative modes of motion in elastic network models [34-36]. Finally, the effect of intrinsically disordered segments may also appear locally. As described in this volume of Current Protein and Peptide Science elsewhere in detail, intrinsic disorder is helpful in i.) increasing the 'capture-radius' of the binding partner followed by a binding-coupled folding event (termed fly-casting mechanism); ii.) increasing the diversity of binding partners; as well as iii.) in providing extra robustness against mutations [37, 38]. Intrinsically disordered regions are, typical to e.g. chaperones, chromatin remodeling proteins or moonlighting proteins [39-41].

**3.B. Mesoscopic disorder of protein structures.** Molten globules characterize the intermediate steps of protein folding, where after the hydrophobic collapse almost all secondary structures of the native state have been formed, but their tertiary structure has not been fully stabilized yet [42-45]. It is an ongoing debate, whether, how and when molten globule structures characterize native proteins in the crowded cellular environment. Recent data show that protein structures may be more flexible, or 'looser' inside the cell than previously thought. It was proposed that an '*in vivo* molten globule structure' may be especially typical to proteins in the vicinity of membranes, where local conditions may stabilize this partially unfolded state [42, 46-49]. We will summarize the protein structural network-related consequences of the molten globule state in the next, 3.C. Section. An interesting example of disordered domains is the tubulin-bound leading head of kinesin, which has a disordered contact structure at its ATP-binding site compared to that of the highly similar trailing head domain. This causes less stable binding of ATP, which makes the leading head dependent on the movement-inducing internal tensions of 12-15 pN in the tubulin-kinesin complex [50]. Intrinsically disordered domains also play a major role in allosteric coupling mechanisms [51].

Order and disorder are important in the eye, which is most probably the single organ of our body where changes in molecular disorder lead to macroscopically *visible* consequences. As a major element of these, the loss of translucency of the eye lens in the aging process leads to the development of cataracts. Fig. **1** shows the protein structure network of a major eye lens protein, human gamma-D-crystallin. The two domains can be easily distinguished as two separate modules of the network. The intermodular region (see dotted circle on Fig. **1**) with sparse network contacts is a typical underconstrained region (see Section 6.B. [22]) that harbors the highly disordered N-terminal and connecting segments of the protein [52]. The loss of the disordered N-terminal arm contributes to an increased aggregation of crystallin with consequent cataract development [53]. The glutamic acid at position 107 (see arrow in Fig. **1**) plays a key role in the association of gamma-crystallin with alpha-crystallin, and cataract formation [54]. As shown in Fig. **1** this key amino acid is adjacent to the disordered region. Loss of disorder at the protein level condenses intra- and inter-protein interactions and contributes to the formation of cataracts. Besides sporadic examples, like that of crystallin, no direct studies have assessed the protein structural network consequences of a large number of intrinsically disordered protein domains.

**3.C. Global disorder of protein structures.** Intrinsically disordered proteins [37, 38] are examples of unfolded states displaying a global disorder under physiologically relevant conditions. However, unfolding is a feature of any protein under relatively extreme conditions. Therefore network studies of the folding process can provide an insight into many naturally occurring scenarios including the molten globule state. The small-worldness of the protein structure network increases during folding as the structure becomes more compact. Proteins with denser network connections (with 'smaller world' than average) are less likely to display a disordered state. Nucleation centres, which were shown to govern folding, are central residues in networks of transient conformations, but often loose their centrality as the protein folds. Similarly, native hubs can be replaced by non-native hubs during unfolding. Hubs which largely preserve their environment during folding/unfolding events are often critical determinants of protein stability [7, 8, 15, 55-58]. A recent study [59] showed that shortcuts in long-range contacts do not lead to faster folding. Thus folding constraints might also explain why protein structure networks are not even smaller worlds. However, detailed studies showing the effect of protein disorder to the small-worldness of protein structure networks are currently missing.

## 4. PROTEIN DYNAMICS-RELATED DISORDER IN NETWORKS

So far, folding has provided an example of dynamic disorder, where the extent of disorder changed over time. In the current Section we extend this view, and treat fluctuations of individual amino acids (Section 4.A.); the effect of surrounding water (Section 4.B.); high mobility of smaller protein segments (Section 4.C.); conformational transitions as sources of dynamic disorder (Section 4.D.). We further analyze the positions of affected residues in protein structure networks.

**4.A. Highly mobile amino acids revealed by elastic network models**. Elastic network models are useful to assess the fluctuation differences between amino acids. The results of network simulations were in good agreement with experimental data on residue flexibility obtained by e.g. X-ray and NMR. Side-chain fluctuations are strongly correlated with the spatial arrangement of the residues, highlighting the fact that central amino acids (occurring on average on more shortest-paths between other residues) display more restricted motion [11]. This is in agreement with the general notion that surface residues undergo larger fluctuations than core residues [25], as well as normal mode analysis that shows that amino acid hubs have lower flexibility than sparsely connected amino acids [18]. Using elastic network model-inferred fluctuations, amino acids could be classified into 3 classes: the highly fluctuating Gly, Ala, Ser, Pro and Asp; intermediate fluctuating amino acids; and weakly fluctuating Ile, Leu, Met, Phe, Tyr, Trp and His [25].

**4.B. Water as a key factor of transitive disorder allowing protein mobility**. Efficient enzyme activity and conformational rearrangements take place through the transient disorder induced by fluctuating water molecules [32]. In a recent summary Hans Frauenfelder and colleagues [60] suggested that "large-scale protein motions are controlled by the solvent viscosity", and "internal protein motions are controlled by hydration". Despite of the importance of water-induced disorder, the dissipative terms arising from water-disorder have been included in the network model representation of protein dynamics only recently [23, 24, 61].

**4.C. Network positions of highly mobile protein segments**. A network-based analysis of local flexibility [22] revealed clusters of flexible amino acids that rigidify upon complex formation. C−H····O-type hydrogen bonds have a major contribution to this dynamic flexibility [62]. PEST-sequences are keys in protein degradation and signaling. PEST-motifs were found to be hyper-flexible and major contributors to protein disorder [63]; however, regretfully a network analysis of flexible clusters and PEST-sequences is missing. Analysis of 103 dynamic α-helices (flipping between helical and coil forms) showed that they are depleted of hydrophobic residues, and have a higher surface accessibility than rigid helices. Helical conformations had more network contacts than the non-helical ones [64].

**4.D. Conformational changes as disorder**. Conformational changes in *cumulus-type* networks have a switch-type energy transfer mechanism (Section 8.B.), while they propagate *via* a multi-trajectory mechanism in *stratus-type* networks (Section 8.B.). Table **1** lists some examples, where a signalling network of amino acids has been identified. A switch-type conformational change is typical in systems where signalling involves a small number of amino acids, like that of subtilisin [24]. This mechanism may be characterized by a 'rigid-body motion', where stiff, hinge-type regions play a decisive role in changing the relative position of the rigid segments [65, 66]. During these changes, *independent dynamic segments* located in the stiffest parts of the protein, and possibly harboring spatially localized vibrations of nonlinear origin such as *discrete breathers*, exchange their energy largely *via* a predominant pathway [24]. In contrast, a multi-trajectory behavior exemplified by the largest, 20 to 50 amino acid clusters of Table **1,** may contain a number of highly populated pathways. Multi-trajectory pathways may converge at conserved inter-modular residues, thus be cross-linked, forming a small-world signalling network [27, 31, 65-71]. A given protein may also involve both mechanisms. In the network analysis of communication processes active centre residues have been shown to possess an exceptionally large centrality. Interestingly, several segments of ligand binding sites were not that central in the information transfer, which may reflect a lower flexibility of binding sites, as opposed to the constraints of the active centre necessary for efficient catalysis [30, 72].

## 5. NETWORK ENTROPY AND PROTEIN STRUCTURE

In this Section first we summarize the current knowledge on general disorder in protein structure networks, as exemplified by the various network entropy functions (Section 5.A.). In the second part (Section 5.B.) we describe attempts to approximate the complex topology of real world networks by an ensemble of simpler networks, such as random or fractal-like network ensembles. Such methods (which resemble the well-known Fourier-transformation of complex functions) may discriminate among ordered, disordered, or intermediary network structures.

5**.A. Network entropy functions as measures of network disorder.** Static entropies characterize the global disorder of network topology. As one of the first approaches to define network entropy, the degree-based entropy is low when the degree of the nodes (number of their neighbors) is uniform [73]. With more complex definitions of entropy which account for the information transfer between all possible pairs of nodes (i.e. all possible clusters of nodes), the analysis of the entropy of representative subnetworks has become equally possible, which helps assessing the entropy of a large network, if we know only a sample of it [74]. This latter approach leads to the methods of Section 5.B., where the entropy of real-world networks, such as protein structure networks, can be characterized by the variability of an ensemble of simpler networks approximating well the original network. The subgraph centrality of E. Estrada [75] may also be used as an entropy function. According to this measure, entropy is smaller for networks that have a single module, as opposed to networks having a well-defined modular structure. Another network entropy function considers the probability distribution of the eigenvalues of the network density matrix [76]. A similar measure based on the spectral moments of the dihedral angle matrix of the protein backbone was used earlier to characterize the degree of protein folding [77]. Entropy functions characterizing dynamical processes, for example a random walk on networks have also been defined. These dynamic entropy functions are useful to characterize the possible speed of spreading processes (like that of epidemics or computer viruses) across networks [78, 79]. Regretfully, a detailed analysis of either static or dynamic network entropy measures of protein structure networks is missing.

**5.B. Network ensembles as entropy measures**. The entropy of complex networks can also be characterized by the variability of a network ensemble approximating well the complex network. The network ensemble may be a group of random networks [80-82] or a series of fractal networks [83]. The size of the automorphism group of the network (obtained by permuting the nodes of the original network preserving their adjacency) also characterizes the entropy of the original network. This automorphism-based entropy is low, if the network is symmetric [84]. Recently the network ensemble method has been developed further to give an entropy-based measure to judge the relevance of network features, such as modularity, or spatial embeddedness [85]. At a coarse-grained level, protein structure networks conform remarkably well to realizations of random networks showing that side-chain interactions exhibit an element of randomness. However, at a finer level, protein structural networks exhibit deviations from randomness [86]. Despite such first approximations, currently network ensemble-based entropy measures are unexplored in protein structure networks.

**6. TYPES OF DISORDER IN NETWORK TOPOLOGY**

In this Section we will summarize our current knowledge of disorder in network topology and the representation of topological network disorder in protein structures. In the strict sense the only 'ordered' network topologies are the entirely regular networks, which do not exist in proteins in their pure form. From this 'purist' point of view, all protein structure networks are disordered. However, there are particular features, which may gauge the level of disorder in these networks. Similarly to our treatment of disorder in protein structures in Section 3., we may discriminate between local (Section 6.A.); mesoscopic (Section 6.B.) and global disorder (Section 6.C.) in protein structure networks. Local network disorder will be exemplified by the level of clustering and assortativity. Mesoscopic network disorder will be characterized by regional deviations from the average link density mostly expressed as the modularity of protein structure networks. Finally, global network disorder will be shown in the examples of network symmetry and topology classes of network phase transitions.

**6.A. Local network disorder: clustering, assortativity and rich-clubs**. Clustering, i.e. the measure of connectivity among a node's neighbors [3, 58] becomes lower, if we take into account only the long-range links of amino acids. This shows that clustering-related network order is mainly introduced by the protein backbone. High clustering usually negatively influences the rate of protein folding [6, 8, 87, 88]. High assortativity refers to preferential attachment of nodes with similar degrees (number of neighbors) and is usually typical of social networks. Protein structure networks display a higher assortativity than other biological networks, which also seems to be caused mainly by the protein backbone. The negative consequence of assortativity in biological networks, i.e. the high vulnerability of the assortative network for 'error-propagation' is helpful in protein structure networks, since from the network standpoint, it may facilitate the perturbation waves of conformational transitions [8, 87, 88]. Hydrophobic hubs of protein structure networks often associate with each other, and form an assortative 'rich club' (Fig. **2A**). Hydrophobic hubs also show a hierarchical 'superhub' structure, which have a large number of hub-neighbors. These structures cannot be observed for the subnetworks of hydrophilic and charged amino acids consistent with the key role of hydrophobic interactions in the core-structure of proteins [1]. Central amino acids do not tend to associate with each other, thus protein structure networks are not 'clumpy' (Fig. **2A**; [89]).

**6.B. Mesoscopic network disorder: link-density differences and modularity**. Disorder can be perceived as a significant deviation from average. In this sense both overconstrained (more crosslinking bonds than average) and underconstrained (less than average) protein regions can be regarded as disordered from the network point of view. Over- and underconstrained regions were shown to correspond to rigid and flexible protein segments, respectively (Fig. **2B**; [22]). A slightly higher global link-density difference occurs in modular structures, where modules are defined as groups of amino acids having a denser link-structure with each other in 3D space than with their neighborhoods. Modules are hierarchical (Fig. **2C**). In this terminology, hierarchy of modules helps protein folding by separating segments of the search for the native conformation to sub-domains of separate '3D link-density groups' [5, 7, 13, 15-18].

**6.C. Global network disorder: symmetry and topology classes of phase transitions**. High network symmetry characterizes a relatively ordered structure, and can be described by multiple hierarchical layers starting from the symmetry of node-pairs, i.e. the assortativity described above. High clustering generally invokes higher-than-average assortativity and modularization [90]. Another form of symmetry, e.g. the symmetry of directed links cannot be directly assessed in protein structure networks, since their links are undirected.

If the resources maintaining network links gradually diminish, networks show rather general topological phase transitions, starting from a random network, and going through network structures with scale-free degree distribution, reaching the star-network, where only a single mega-hub exists and finally disassembling to smaller sub-networks (Fig. **3**; [91, 92]). With their degree-limited, quasi scale-free degree distribution, protein structures emerge as rather disordered systems.

**7. TYPES OF DISORDER IN NETWORK DYNAMICS**

In this Section we will highlight a few major types of disorder in network representations of protein structure and dynamics. We will start with the propagation of perturbations (Section 7.A.), and continue with possible applications of the methods to describe the synchronicity of network oscillators to the vibrations of individual amino acids (Section 7.B.). We will conclude this Section by describing independent dynamic segments that harbor discrete breathers, and as representing the general category of creative elements (Section 7.C.). We will show that this concluding example of dynamic network disorder is a key factor to understanding the mechanism of many conformational changes, which were summarized in Section 4.

**7.A. Propagation of perturbations in protein structure networks**. Perturbations are the most common dynamical disorder in protein structure networks, since they change the 'order' of the resting state. The propagation of perturbations in both protein structure and protein-protein interaction networks has been reviewed recently by Antal et al. [93]. They suggested a set of conditions for modeling the phenomenon. The small-world property of protein structure networks dampens fluctuations [94]. An earlier study examined the effect of an external harmonic perturbation applied to one of the synchronized phase-coupled oscillators in a random network, and showed that the dissipation is close to exponential [95]. Functionally active residues were found to possess enhanced communication propensities in a Markov-model of perturbation propagation [30]. The perturbation response scanning technique of Atilgan et al. [96] revealed that binding-induced conformational change might be achieved through the perturbation of residues scattered throughout a subset of proteins, whereas in other proteins the perturbation of residues remains confined to a highly specific region [96]. Besides these studies, most of the other examples in Section 4.D. and Table **1** were also based on the analysis of perturbation propagation in protein structure or related networks. However, more general studies on the effects of disorder on the propagation of perturbation are currently missing. A general Matlab-compatible program to describe network perturbations, called Turbine, is in construction in the authors' laboratory [97], and will be freely available during 2011 on the website, www.linkgroup.hu. This tool or the application of the PerturbationAnalyzer program of protein-protein interaction networks [98] to protein structure networks may help perturbation studies of protein structure networks in the future.

**7.B. Synchronicity of oscillator networks**. 'Order' in the dynamics of protein structure networks is described by the synchrony of amino acid-related oscillators. Small-world property, which is typical to protein structures, enhances synchronization [94]. Estrada and Hatano [75] found that amino acid oscillators of protein structure networks display a multiple synchronization pattern, and may fluctuate between various states. This finding demonstrates the importance of disorder in oscillation synchronicity, and is in agreement with the multiple conformational states of many proteins. According to a recent study, oscillations of protein structure networks fall into the stable region of oscillation patterns. The small-world character seems to be a main reason for the stability of oscillations in the 56 protein structures examined [99]. Both coupling-delays and coupling strength of the oscillators affect the stability of the collective behavior [100]. In agreement with this, a model of a G-protein coupled receptor using a network of fluctuating harmonic oscillators and discriminating α-helices and loops with different spring elastic constants showed that this model system described real situations much better, once the discrimination between α-helices and loops was applied [101]. Many of the elastic network models in Section 2 accommodate the dynamical differences of protein segments in greater detail.

**7.C. Independent dynamic segments hosting discrete breather vibrations and creative elements**. As shown in the preceding paragraphs, disorder in the dynamics is not uniform across all segments of the protein structure. Key segments of conformational transition are primary candidates for extreme dynamical disorder (i.e. extreme individuality in their dynamics). The studies of Piazza and Sanejouand [23, 24] employed a nonlinear elastic network model, where the effect of surrounding water molecules was also taken into account as a dissipation term added to residues on the protein surface. These studies revealed the existence of 'discrete breathers', i.e. extremely individual spatially localized vibration patterns lying within short protein segments at very specific locations. The regions harboring discrete breathers were termed later as independent dynamic segments [31]. These segments are located usually at the stiffest regions of the protein, and often co-localize with active centers.

Discrete breathers on the independent dynamic segments are reminiscent of Davydov-solitons [102], even if they are not solitons in the exact sense. The original concept of Davydov [102] involved α-helices. However, the soliton concept was extended to many types of protein structures [103-106], and was also linked to low-frequency phonons (vibrations) [107]. Discrete breathers show a different dynamic behavior than other segments of the protein structure network, and therefore they are key elements of disorder in network dynamics. In model systems structural disorder was shown to facilitate transmission of solitons [108]. Furthermore, discrete breathers were shown to be able to mediate long-range energy transfer under quite general conditions. Localized perturbations in some proteins, such as kinetic energy kicks, turned out to excite discrete breathers at distant locations, acting as extremely efficient and irreversible energy harvesting centers [23, 24]. The comparative analysis of protein disorder, soliton and discrete breather topology and dynamics will be an exciting field of future studies.

Independent dynamic segments often occupy a more locally central position in the sense that their neighbors in the protein structure networks are not connected with each other. This position and the extremity in network dynamics is typical to the creative elements defined by Csermely in 2008 [109]. Creative elements are the least predictable, highly mobile elements occupying various locally and globally central positions in the network [109]. Thus creative elements are a major source of disorder in network dynamics. How large is the overlap between hinges, independent dynamic segments (discrete breathers), creative elements and inter-modular connecting nodes awaits future studies.

**8. EFFECTS OF NETWORK DISORDERS ON PROTEIN FUNCTION**

In this Section we will summarize the effects of structural and dynamical disorder of networks on protein function. We warn the reader that this Section will be – in part – speculative, since there is little data to link these different approaches of network and protein science. First, we will describe the potential connections between disorder, flexibility and conformational changes introducing the 'rigidity pathway' and 'frustration tube' concepts (Section 8.A.). Next, we will describe a hypothetical classification of proteins according to their global network structure, which is connected to their rigidity pattern as well as the mechanism of their conformational changes (Section 8.B.). We will show that disorder helps proteins to visit hidden, rare conformational states (Section 8.C.) and that at the level of game theory disorder may also help and not only prevent cooperative behavior (Section 8.D.). We will conclude the Section with the hypotheses that disordered protein regions may act as a driving force for the evolution of other protein segments helping to reach 'rarely visited' genotypes, and that evolution of protein structures can be perceived as a generation of a dynamic disorder constrained by the features of both protein structure and higher level networks (Section 8.E.).

**8.A. Connections between disorder, flexibility and conformational change.** As we described in Sections 3. and 4., local and global disorder cause increased flexibility. Importantly, a general increase in flexibility also highlights the remaining, rigid parts of proteins. These stiff parts often coincide with the hinges and the independent dynamic regions described in Sections 4.D. and 7.C. Using the network-related, mainly local flexibility and rigidity definition of Jacobs et al. [22] described in Section 6.B., Rader and Brown [110] showed that two thirds of the 16 examined proteins had a rigid path spanning 2 to 4 nm connecting the effector and catalytic sites. It is an open question how general is the coincidence between rigid regions and connection pathways. Disorder of residues participating in intra-protein signaling pathways is described as 'frustration' by Zhuravlev and Papoian [28], meaning a predisposition to change, which is usually 'flipping' of pathway residues upon ligand binding. Residues on evolutionarily conserved paths usually reside in the high-density core of the protein, and thus have a high burial free energy, which makes their evolutionary change energetically expensive. Highly frustrated residues may act in coordination forming a propagating 'frustration front' from the effector site towards the catalytic site in a 'frustration tube' [28]. The 'rigidity propagation model' of Jacobs et al. [111] was originally described using peptide toy models and the α-helix/coil transition. A hypothetical extension of the model may invoke the propagation of rigidity with the propagation of the 'frustration front'. Such a mechanism would use a combination of the rigidity path concept of Rader and Brown [110] with the 'frustration front' concept of Zhuravlev and Papoian [28]. It may also be true, however, that the 'rigidity path' and the 'frustration front' may characterize distinct classes of proteins, like those two described in the following Section, 8.B.

**8.B. Hypothetical classification of proteins to 'cumulus' and 'stratus' classes.** The perturbation studies of Atilgan et al. [96] described in Section 7.A. showed that proteins might be classified in two classes: 'class I. proteins' can be studied well with the normal mode analysis of elastic network models, while this method does not give such straightforward results for 'class III. proteins'. Table **2** lists other potential properties of these protein classes. The features of class I. and class III. proteins fit well the duality of 'cumulus-type' and 'stratus-type' networks, respectively. Networks with a cumulus topology [112] have a rather disjoint, multi-centered modular structure resembling that of the altocumulus clouds. Such a structure has a rather limited overlap between the modules [113], which implies higher rigidity of the individual modules. Networks with a stratus topology [112] have a rather condensed, compact structure resembling that of stratus clouds. Such networks have a significant overlap between their modules ([113], Fig. **4**), which implies higher flexibility of the entire network. This protein classification agrees well with the previously observed duality in the mechanism of conformational changes.

Rigid-body motion, where stiff, conserved, hinge-type regions play a decisive role in changing the relative position of the rigid segments [65, 66, 114] may be typical for cumulus-type proteins. Cumulus-type proteins may also harbor independent dynamic segments behaving as discrete breathers, and able to transmit conformational changes *via* energy transfer [23, 24, 31]. The Davydov-soliton theory [102] and its applications [103, 104, 107] described in Section 7.C. fit much better cumulus- than stratus-type proteins.

Proteins with a transition from cumulus-type to stratus-type protein structure network topology display redundant pathways converging at modular boundaries [27, 31, 65-71] as described in Sections 4.D. and 7.C. (Such proteins may correspond to the intermediate, class II. group of Atilgan et al. [96].)

Stress induces a stratus-like → cumulus transition in yeast protein-protein interaction networks. This transition seems to be a general phenomenon, which suggests that protein structure networks may undergo a similar, stratus-like → cumulus transition upon mechanical stress, or upon a high demand to increase their efficiency [113, 115].

**8.C. The help of network disorders to visit rare, hidden conformations**. Disorder in both network topology and dynamics – exemplified by creative elements [109] – may help a protein to reach rarely visited, 'hidden' conformations, as have been recently uncovered in the enzymatic action of the human proline isomerase, cyclophilin A, or in the phosphorylation (i.e. signal transduction) of the nitrogen regulatory protein C signaling protein [116, 117]. These extreme, minor conformational states represent a stability point, which is distant from the frequently visited members of the conformational ensemble at the conformational energy landscape. The long-range jumps required to reach these states often require a significant disorder [31, 32].

**8.D. Disorder and cooperation: spatial games of amino acids**. Cooperativity is a cornerstone of protein folding, conformational changes and enzyme function. Local cooperativity exemplified by the formation of an α-helix is disturbed by protein disorder. The effect of disorder on global cooperativity, however, cannot be assessed by simple assumptions and terms. Game theory is a good tool to study the evolution of cooperativity of individual agents. Recent advances in spatial games allow studies of the cooperation of several agents, like proteins in complex networks [31-32, 118, 119]. Such studies may be extended to protein structure networks, where agents participating in the game will be the amino acids of the protein. Dynamic disorder of proteins may be modeled by changes of the strategy update rules of the amino acids. Low probability random changes of strategy update rules significantly increased cooperation [120], which shows that at the global level, dynamic disorder may, in fact, help and not only prevent cooperative behavior.

**8.E. Evolution of proteins in terms of dynamic disorder in the 'looooong' run**. Disordered protein regions ease the evolutionary constraints, and may act as a driving force for the evolution of other protein segments. However, some amino acids involved in forming protein-protein interfaces, like tryptophan and tyrosine remain rather conserved even in disordered proteins [121-124]. Showing the important function of disorder, functionally important disordered regions can be conserved in evolution [125-127]. Disordered protein regions may drive the evolution of proteins towards similar, 'rarely visited' phenotypes like the hidden conformations described in Section 8.C. This assumption is supported by recent data showing that evolving systems with 'hidden', rarely visited phenotypes need an intermediate level of robustness for fast adaptation [128]. Intermediate robustness is served well by an intermediate level of disorder. On its own, evolution of protein structures can be perceived as generation of dynamic disorder from the original 'order' of the protein structure by mutations. Links between disorder and the recently described protein sectors having a similar evolutionary history [129] provide a promising field to study. Similarly, network studies may also reveal how the mutation-induced disorder is tolerated by binding partners, in the higher level network of protein-protein interactions [130]. The effect of disorder on networks of higher cellular hierarchy, such as protein-protein interaction networks (interactomes), or signaling networks (signalomes) will be the subject of the next Section, Section 9.

## 9. EFFECTS OF INTRINSICALLY DISORDERED PROTEINS ON INTERACTOMES, SIGNALOMES AND OTHER CELLULAR NETWORKS

Just as the nodes in a network are not independent from each other, the cellular networks at different hierarchical levels are also not independent. In these hierarchically embedded, interdependent networks disorders appearing in one network could easily propagate to others and thus the well-known error-tolerance property of scale-free networks may become significantly compromised [131, 132]. An illustrative hierarchy of networks in the cell nucleus starting from protein structure networks and ending at the chromatin network is shown on Fig. **5.** In the following sub-Sections, we will discuss how disorder in the bottom-level networks of protein structures could perturb higher level biological networks, such as protein-protein interaction, or signaling networks. We will also describe the effects of aging to induce a propagating disorder in the hierarchically embedded cellular networks.

**9.A. Propagation of protein disorder to the structure of higher level networks.** Accumulating evidence suggests that disorder in protein structures is directly linked to protein-protein interaction, signaling, transcriptional regulatory and possibly to chromatin interaction networks [39, 133-135]. Date-hubs are nodes of protein-protein interaction networks changing their neighbors quite often. They are enriched in intrinsically disordered proteins [37, 134-136], which reflect the versatile binding character of intrinsically disordered proteins at a higher network representation. Intrinsically disordered proteins often behave as moonlighting proteins [41], which shows another aspect of their pleiotropic binding character. Intrinsically disordered, non-hub proteins tend to bind each other in protein-protein interaction networks. This feature is also true for disordered signaling proteins [137].

Signaling networks are enriched in disordered proteins [138]. Intrinsic disorder in protein structures enhances and extends their interactions by optimizing intra-molecular site-to-site allosteric coupling and inter-molecular oligomerization [51, 139], as well as by the pleiotropic binding pattern described before. Intrinsically disordered proteins have typically larger and tighter interfaces, which help to stabilize their complexes [140, 141]. As a special example of the pleiotropic binding pattern, it was suggested that disorder in a transcription factor positively correlates with the number of targets the actual factor may bind to [134]. Transcription factor domains and their cofactors, which could modulate the networks of *cis* and *trans* chromatin interactions, also harbor a significant intrinsic disorder [39], with tumor suppressors as well known examples. Alterations in protein structure networks of disordered proteins may significantly enrich the complexity of higher order networks.

Under the *protein trinity* paradigm [142], single nucleotide polymorphisms (SNPs) can be related to the activity and stability of disordered proteins. Certain SNPs could affect the intra- and inter-molecular site recognition mediated by protein disorder, by shifting the landscape which might alter their multi-factorial binding to other moieties in the cell. SNPs in disordered segments, like in PEST segments [63, 143], which are the prime targets for protein degradation and post-translational modifications, were proposed to alter the stability and dynamics of protein structure networks [144]. Increase or decrease in the dynamics and stability of protein structure [145], could alter the dynamics of interactomes and signalomes by facilitating promiscuous interactions [146], or by limiting and disproportioning the key interactions. Similarly, SNP-induced, mutation-dependent, or other disorders in protein structure could propagate to gene-regulatory and chromatin interaction networks either directly from protein structure networks of disordered transcription factors and chromatin architecture proteins, or indirectly, *via* their neighborhood in protein-protein interaction networks.

**9.B**. **Effects of protein disorder on the function of higher level networks.** Intrinsically disordered proteins are strongly associated with cancer and other disease phenotypes [138, 147-149]. Intrinsic disorder is also tightly linked to the recognition of misfolded proteins playing a prominent role in a number of diseases such as in neurodegeneration. Molecular chaperones recognizing misfolded proteins are enriched in disordered regions [40] and misfolded substrates are bound by disordered regions of ubiquitin-ligase [150] directing them to proteasomal degradation. A recent study showed that disordered proteins are preferentially sequestered by amyloid aggregates, the hallmarks of neurodegenerative diseases, such as Alzheimer's and Parkinson's diseases [151]. It is possible that errors in the stability and dynamics of disordered proteins perturb the higher level networks in a pleiotropic manner. Thus, multiple errors in distinct protein networks could create mayhem of dysregulations leading to complex pathological conditions.

**9.C. Aging, as a special cause of propagating disorder.** Aging induces a number of chemical changes of amino acids including oxidation, deamidation of asparaginyl and glutaminyl residues and the subsequent formation of isopeptide bonds, glycation, etc. [152, 153]. These changes lead to a significant structural disorder and misfolding-induced protein aggregation overloading both the chaperone and proteasomal degradation machineries and posing a great danger to the aging cell [154, 155]. Misfolding-induced disorder at the protein level significantly contributes to the disorganization of aging networks and to the increased noise of cellular processes in aging organisms [156].

**10. CONCLUSIONS AND PERSPECTIVES**

In this review we detailed the importance of analyzing protein dynamics and structure in terms of associated networks to understand the effects of disorder on protein and cellular function. We showed how network disorder is reflected in protein structure and dynamics. Now we will highlight the major points, where we predict progress on this rapidly expanding field:

- systematic studies on the position of glycines, prolines and loops in protein structure networks;
- network studies of intrinsically disordered segments, domains and proteins; links between disorder and the small worldness of protein networks;
- network analysis of locally dynamic protein segments, such as flexible amino acid clusters, PEST-sequences, etc.;
- assortativity of modules of protein structure networks;
- entropy measures of protein structure networks;
- the role of structural disorder in perturbation propagation in protein structure networks;
- the application of spatial games to the amino acids as playing agents;
- the relationship between disordered protein segments and protein sectors;
- elucidation of protein structure network pathways transmitting conformational transitions;
- the extent of overlap between inter-modular connecting nodes, hinges, independent dynamic segments harboring discrete breathers, solitons and creative elements;
- the significance of rigid regions in the transmission of conformational changes;
- effects of protein disorder on conformational and evolutionary networks; and finally
- effects of protein disorder on the functions of cellular networks.

We are at the very beginning of the understanding of the interrelationship of protein and network disorder, as well as of the elucidation of their effects on the functions of proteins and cells. However, from the initial studies summarized in our paper, it is already clear that this field will be rich in surprises in the coming years, and will teach us a lot on the organization of living matter.

**ACKNOWLEDGEMENTS**

Authors would like to thank members of the LINK-group (www.linkgroup.hu) especially Mr. Ágoston Mihalik for helpful suggestions. F.P. would like to thank Y.-H. Sanejouand and P. De Los Rios for enlightening discussions. Work in the authors' laboratory was supported by research grants from the Hungarian National Science Foundation (OTKA-K69105 and OTKA-K83314) and from the EU (FP6-016003). This project has been funded, in part, with federal funds from the NCI, NIH, under contract HHSN261200800001E. This research was supported, in part, by the Intramural Research Program of the NIH, National Cancer Institute, Center for Cancer Research. The content of this publication does not necessarily reflect the views or policies of the Department of Health and Human Services, nor does mention of trade names, commercial products, or organizations imply endorsement by the U.S. Government.

**Table 1. Examples of dynamic disorder: network-identified signalling pathways in protein structures**

| Name of protein | Number of amino acids involved | Method(s) of detection | Function of pathway | References |
|---|---|---|---|---|
| subtilisin | 3 | nonlinear elastic network model | transfer of energy from Leu-42 via Ala-85 to Val-177 2.3 nm apart | [Piazza and Sanejouand, 2009] |
| phosphofructo-kinase, lactate dehydrogenase | 6 to 8 | detection of contact rearrangement network | signalling between allosteric effector and active site | [Daily et al., 2008] |
| subtilisin inhibitor, CI2 | 8 | analysis of network shortest paths | propagation of subtilisin binding-induced conformational change | [Atilgan et al., 2004] |
| GroEL chaperone | 15 to 20 | mutational studies, structural perturbation and information propagation of an elastic network model | propagation of ATP-binding and hydrolysis-induced conformational changes in GroEL and between GroEL and GroES | [Chennubhotla and Bahar, 2006; Kass and Horovitz, 2002; Tehver et al., 2009] |
| tRNA-synthetases | 50 | molecular dynamics simulations and network analysis | conformational coupling between tRNA and the activated methionine | [Gosh and Vishveshwara, 2007; Sethi et al., 2009] |

**Table 2. Features of proteins having 'cumulus-type' or 'stratus-type' protein structure networks**

| Name of character | Features of cumulus-type proteins | Features of stratus-type proteins |
|---|---|---|
| global topology and dynamics of protein structure network | the cumulus-type protein structure network has many distinct, relatively non-overlapping modules (the dynamics of the whole network, i.e. the whole protein can be simplified to that of a few interacting segments) | the stratus-type protein structure network has a highly overlapping, not so well defined modular structure (the protein displays a highly complex dynamics, which can not be simplified to the interaction of a low number of segments) |
| number of modes describing the conformational change in normal mode analysis | a single, or very few collective modes are enough to describe the conformational change (class I. proteins of Atilgan et al. [Atilgan et al., 2010b]) | many modes are needed to describe the conformational change (class III. proteins of Atilgan et al. [Atilgan et al., 2010b]) |
| predicted mechanism of conformational transfer | very fast, very efficient, and highly directed (predictable) energy transfer between a very few selected sites | the conformational change spreads over a multitude of intra-modular pathways (at its extreme almost resembling a random walk mechanism)* |
| segments playing a key role in conformational changes | hinges, individual dynamic segments behaving as discrete breathers (and/or solitons) | in the extreme form of a stratus-type protein structure network, all amino acids participate almost equally in the signal transmission of conformational changes* |

*In the reality the two classes of the cumulus-type and stratus-type proteins of the Table are not that separated and intermediate protein structure network topologies also exist as shown by the intermediate, class II. proteins of Atilgan et al. [Atilgan et al., 2010b]. In such intermediate cases inter-modular, conserved amino acids emerge, where the multiple, redundant pathways converge. If the topology of the protein structure network shifts close to that of a cumulus-type network, these inter-modular key amino acids may start to resemble the hinges and individual dynamic segments (discrete breathers, solitons), and start to behave as sources and sinks of an energy transfer mechanism.

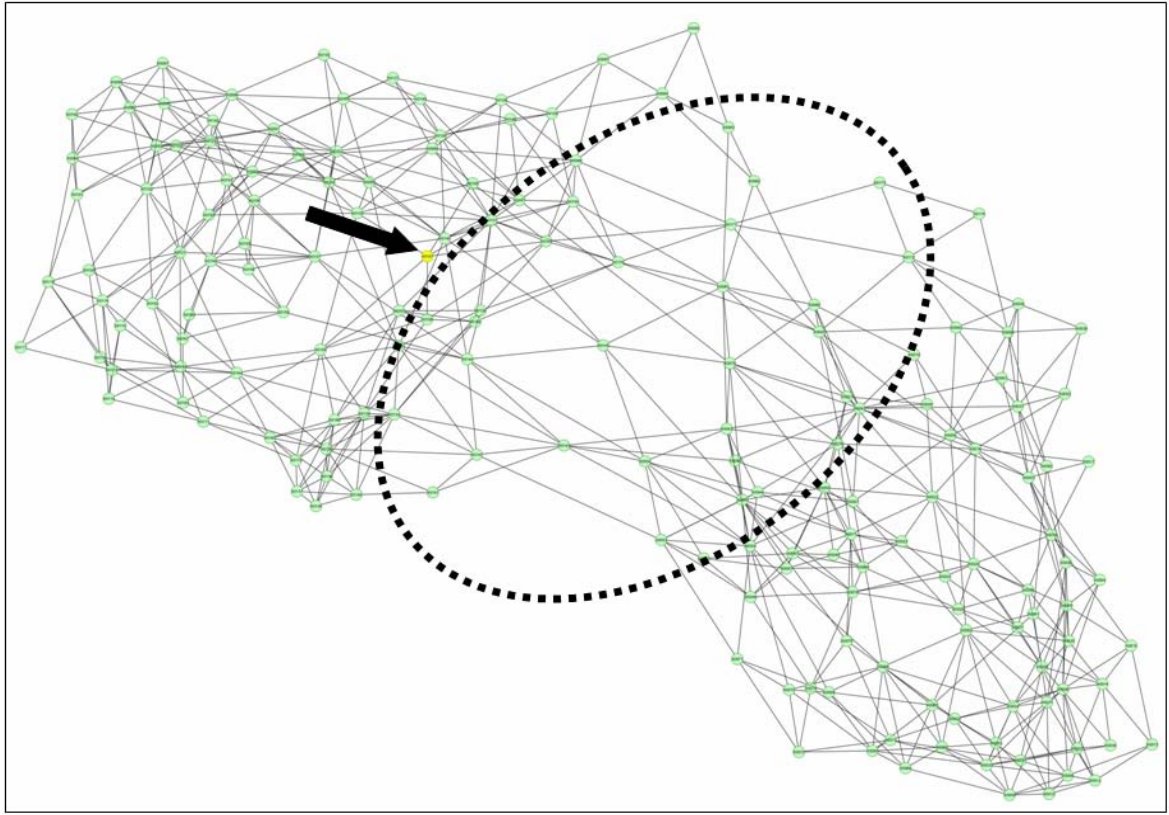

**Fig. (1).** Gamma-crystallin: an illustrative example of the key importance of disordered regions in protein function. The protein structure network of human gamma-D-crystallin is shown (pdb code: 1hk0) containing all non-covalent amino-acid contacts within a threshold of 0.4 nm was determined by the HBPlus program [McDonald et al., 1994]. The protein structure network was visualized using the Cytoscape program [Kohl et al., 2011]. The central region highlighted with the dotted circle contains the highly disordered N-terminal and connecting segments of gamma-crystallin [Wu et al., 2005]. The loss of the disordered N-terminal arm contributes to an increased aggregation of various forms of crystallin leading to the development of cataract in aging eye lenses [Robertson et al., 2008]. The arrow points to the adjacent 107-Glu playing a key role in cataract formation [Banerjee et al., 2011].

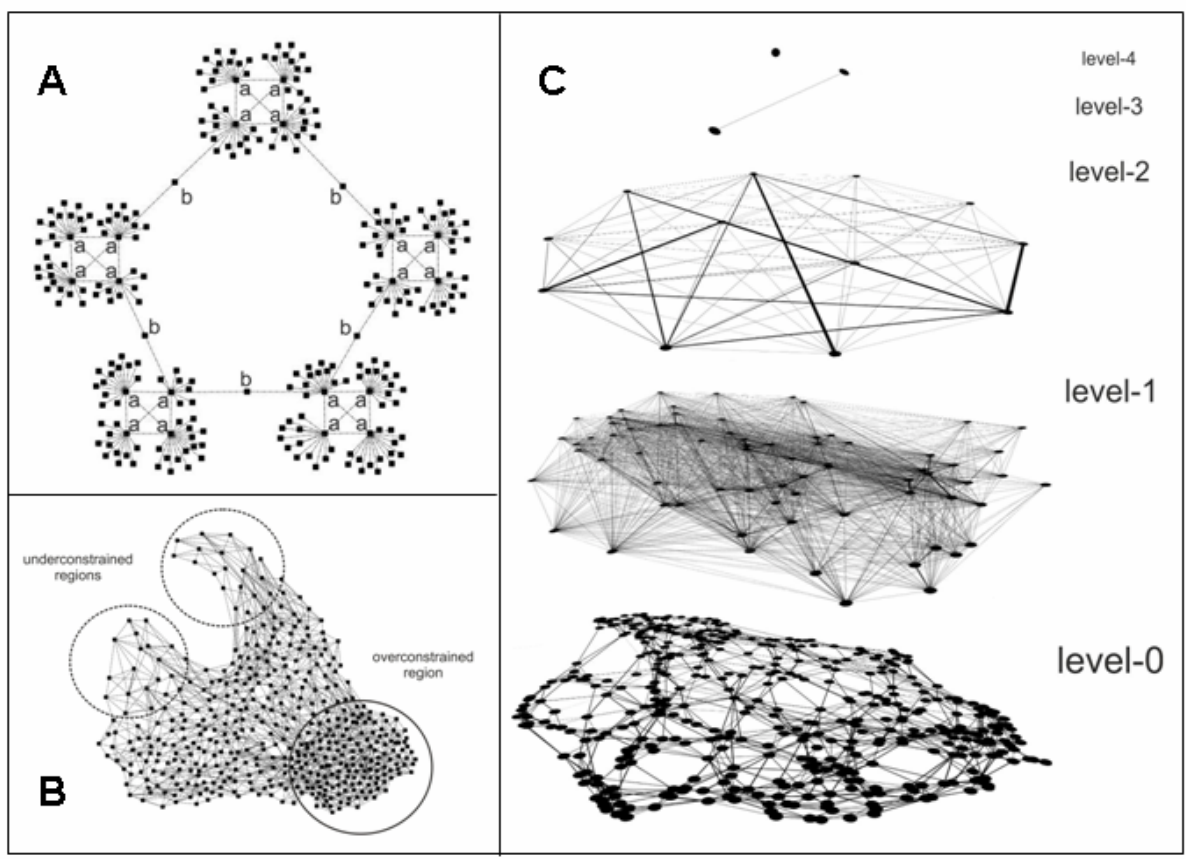

**Fig. (2).** Types of network disorder and their possible effects on protein function. This illustrative figure shows three different topologies affecting network disorder. Panel A: Hydrophobic amino acid networks are 'rich clubs', which infers that their hubs (marked with "a"-s) are preferentially connected with each other [Aftabuddin and Kundu, 2007]. The illustrative network also shows that protein structure networks are not 'clumpy', which means that globally central amino acids (marked with "b"-s) are not connected to each other [Estrada et al., 2008]. Panel B: Illustrative under- and overconstrained regions of protein structure networks corresponding to flexible and rigid regions, respectively [Jacobs et al., 2001]. Flexible regions may contain intrinsically disordered segments. Panel C: Modular hierarchy of an illustrative protein structure network (level 0: the original network; level 1: network modules; level 2: modules of modules; level 3: third generation super-modules; level 4: at this level the whole network finally coalesced into a single element).

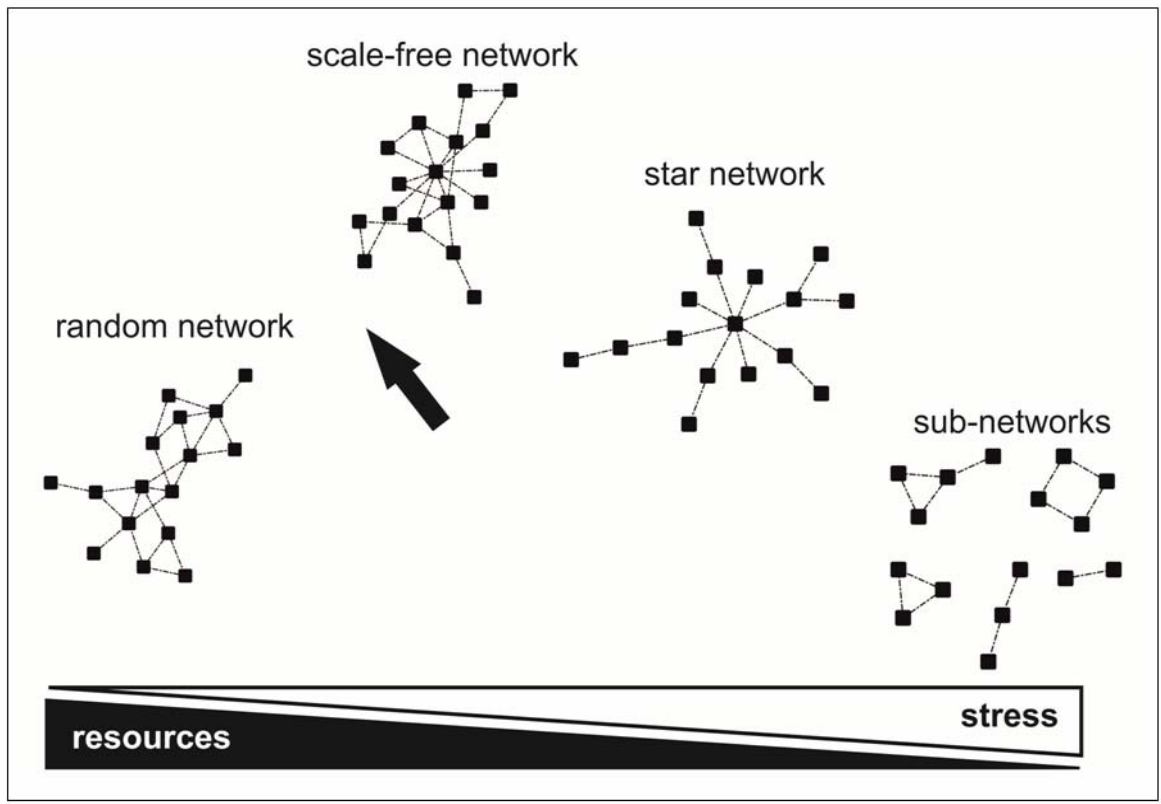

**Fig. (3).** The position of protein structure networks in the series of network phase transitions shows their general disorder. If resources diminish and/or stress increases, networks undergo topological phase transitions. The figure illustrates these transitions starting from a random network, passing through scale-free and star-networks, and finally disassembling to smaller sub-networks [Csermely, 2006; Derényi et al., 2004]. The arrow points to the position of protein structure networks, which are rather disordered with their degree-limited, quasi scale-free degree distribution, if compared to star networks or sub-networks.

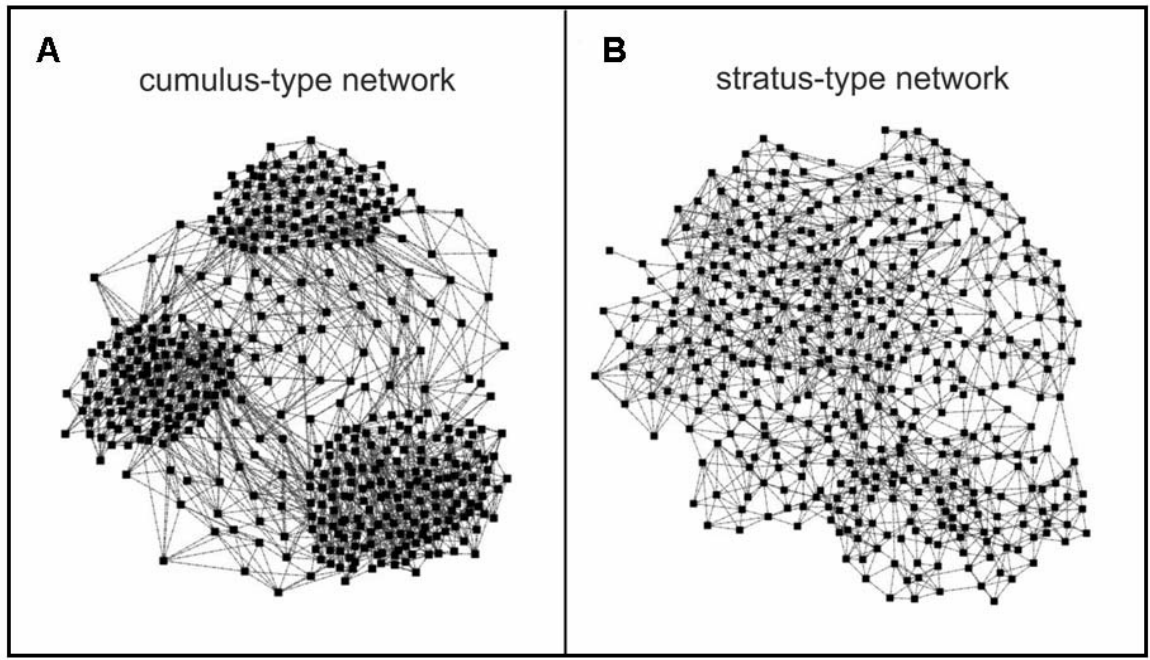

**Fig. (4).** Cumulus- and stratus-type networks. This illustrative figure compares the topologies of cumulus-type (Panel A) and stratus-type networks (Panel B), where the cumulus-type protein structure network has many distinct, relatively non-overlapping modules, as opposed to the stratus-type protein structure network, which has a highly overlapping, not so well defined modular structure [Batada et al., 2006; Mihalik et al., 2008]. In the reality mixtures of the two, rather extreme types shown on the Figure also exist.

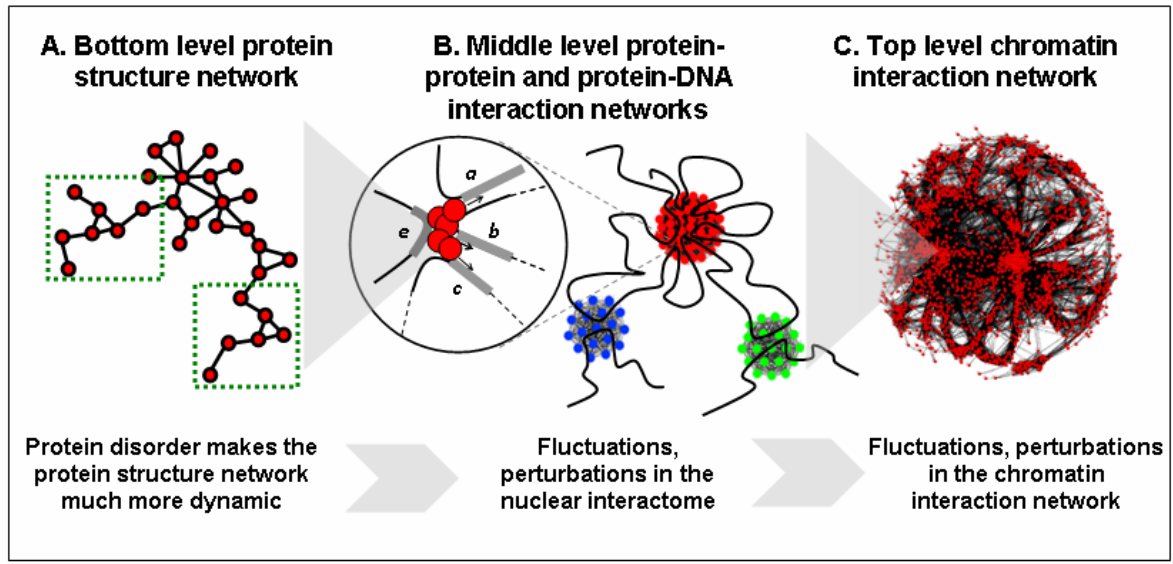

**Fig. (5).** Dynamical perturbation propagation in interdependent, hierarchical networks. The figure illustrates the propagation of protein disorder to higher level networks in the example of the cell-nucleus. Protein disorder makes the bottom level protein structure network much more dynamic, which causes fluctuations in the top level chromatin interaction network *via* perturbations of protein-protein and protein-DNA interaction networks. Panel A. An illustrative amino acid network of a disordered protein. Dots represent amino acids and links stand for their physical interactions. The highly dynamic, disordered regions are marked with dotted boxes. Panel B. Illustrative figure of spherical representation of nuclear protein-protein and protein-DNA interaction networks. Letters *a, b* and *c* denote gene loci, while *e* signifies a distal regulatory element, e.g. an enhancer. The three protein complexes depict the protein-protein interaction networks involved in distinct nuclear functions like transcription, replication, DNA-repair, etc. Panel C. Human chromatin interaction network constructed from Hi-C data [Lieberman-Aiden et al., 2009]. Nodes mark distinct chromatin domains of 1 megabase length.